\documentclass[11pt,a4paper]{article}
\usepackage{jcappub}

\usepackage[latin1]{inputenc}
\usepackage{graphicx}
\usepackage{epsfig}
\usepackage{color}
\usepackage{comment}
\usepackage{slashed}

\title{Limits on the Neutrino Velocity, Lorentz Invariance, and the Weak Equivalence Principle with TeV Neutrinos
from Gamma-Ray Bursts}

\author[a,b]{Jun-Jie Wei,}
\author[a,c]{Xue-Feng Wu,}
\author[d]{He Gao}
\author[e,f,g]{and Peter M{\'e}sz{\'a}ros}

\affiliation[a]{Purple Mountain Observatory, Chinese Academy of Sciences, Nanjing 210008, China}
\affiliation[b]{Guangxi Key Laboratory for Relativistic Astrophysics, Nanning 530004, China}
\affiliation[c]{Joint Center for Particle, Nuclear Physics and Cosmology, Nanjing
University-Purple Mountain Observatory, Nanjing 210008, China}
\affiliation[d]{Department of Astronomy, Beijing Normal University, Beijing 100875, China}
\affiliation[e]{Department of Astronomy and Astrophysics, Pennsylvania State University, 525 Davey Laboratory, University Park, PA 16802}
\affiliation[f]{Department of Physics, Pennsylvania State University, 104 Davey Laboratory, University Park, PA 16802}
\affiliation[g]{Center for Particle and Gravitational Astrophysics, Institute for Gravitation and the Cosmos, Pennsylvania State University, 525 Davey
Laboratory, University Park, PA 16802}

\emailAdd{jjwei@pmo.ac.cn; xfwu@pmo.ac.cn; gaohe@bnu.edu.cn; nnp@psu.edu}

\abstract{Five TeV neutrino events weakly correlated with five gamma-ray bursts (GRBs) were detected recently by
IceCube. This work is an attempt to show that if the GRB identifications are verified, the observed time delays between the TeV neutrinos
and gamma-ray photons from GRBs provide attractive candidates for testing fundamental physics with high
accuracy. Based on the assumed associations between the TeV neutrinos and GRBs, we find that
the limiting velocity of the neutrinos is equal to that of photons to an accuracy of
$\sim1.9\times10^{-15}-2.5\times10^{-18}$, which is about $10^{4}-10^{7}$ times better than the
constraint obtained with the neutrino possibly from a blazar flare. In addition, we set the most stringent limits
up to date on the energy scale of quantum gravity for both the linear and quadratic violations of
Lorentz invariance, namely $E_{\rm QG, 1}>6.3\times10^{18}-1.5\times10^{21}$ GeV and
$E_{\rm QG, 2}>2.0\times10^{11}-4.2\times10^{12}$ GeV, which are essentially as good as or are an improvement
of one order of magnitude over the results previously obtained by the GeV photons of GRB 090510 and
the PeV neutrino from a blazar flare. Assuming that the Shapiro time delay is caused by
the gravitational potential of the Laniakea supercluster of galaxies, we also place the
tightest limits to date on Einstein's weak equivalence principle through the relative
differential variations of the parameterized post-Newtonian parameter
$\gamma$ values for two different species of particles (i.e., neutrinos and photons),
yielding $\Delta\gamma \sim 10^{-11}-10^{-13}$.
However, it should be emphasized again that these limits here obtained are
at best forecast of what could be achieved if the GRB/neutrino correlations would be finally confirmed.}


\begin{document}
\maketitle

 \flushbottom


\section{Introduction}
The IceCube Collaboration recently performed a search for neutrinos of all flavors during the prompt
emission of 807 gamma-ray bursts (GRBs) over the entire sky \cite{IceCube Collaboration et al.2016}.
This search in three years of IceCube data resulted in five cascade neutrino candidate events
correlated with five GRBs.
These spatial and temporal coincidences yielded a combined P-value of 0.32,
which was obtained from the total test statistic value with respect to the background-only distribution.
In other words, the significance of these neutrino events being associated with GRBs is not high,
and they are compatible with the background expectation.
Nevertheless, it has been proposed that such associations between the high energy cosmic neutrinos and GRBs,
if truly confirmed in the future, would be a powerful tool to constrain violations of the Lorentz invariance
and equivalence principle (see e.g. \cite{Jacob et al.2007,Gao et al.2015,Wang et al.2016}).
We now discuss the significant capability of these five associations for testing fundamental physics,
such as the neutrino velocity, Lorentz invariance, and the weak equivalence principle (WEP).

Recently, a theoretical constraint on superluminal neutrino velocity, $|v-c|/c\leq(0.5-1.0)\times10^{-20}$,
has been obtained by treating kinematically allowed energy losses of superluminal neutrinos arising from
vacuum pair emission and neutrino splitting \cite{Stecker et al.2015}. The nearly simultaneous detection of
neutrinos and light from astronomical sources provide a useful model-independent method for constraining
the neutrino velocity. Using the arrival time delay of neutrinos and photons from supernova 1987A,
Ref. \cite{Longo1987} showed that the velocity of neutrinos is equal to that of photons to an accuracy of
$\sim2\times10^{-9}$.  This result was further improved by \cite{Kadler et al.2016} who used the same method
and placed a limit on the relative velocity difference of $|v-c|/c \lesssim {\mathcal O}(10^{-11})$ from
an association between a PeV neutrino and the outburst activity of blazar PKS B1424-418.

Lorentz invariance is one of the fundamental principles of special relativity.
However, some Quantum Gravity (QG) models predict that quantum fluctuations in the space-time
would make it appear `foamy' on short time and distance scales \cite{Amelino-Camelia et al.1997}.
Thus, the propagation of massless particles (photons, or in the limit, neutrinos) through the
foamy structure of space-time would exhibit a non-trivial dispersion relation in vacuo
\cite{Amelino-Camelia et al.1998} which could lead to the violation of Lorentz invariance (LIV).
Since it is typically expected for QG to manifest itself fully at the Planck scale,
the QG energy scale $E_{\rm QG}$ is approximated to the Planck energy scale $E_{\rm Pl}$, i.e.,
$E_{\rm QG}\approx E_{\rm Pl}=\sqrt{\hbar c^{5}/G}\simeq1.22\times10^{19}$ GeV
(see \cite{Mattingly2005,Amelino-Camelia2013}, and references therein). Hence,
$E_{\rm Pl}$ is a natural scale at which one expects Lorentz invariance to be broken.
As a result of LIV effect, the speed of propagation of particles (neutrinos or photons) could become
energy-dependent in vacuum (see e.g. \cite{Amelino-Camelia et al.1998,Choubey et al.2004,Ellis et al.2013}).
The leading term in the modified dispersion relation is
\begin{equation}
E^{2}\simeq p^{2}c^{2}+m^{2}c^{4} \pm E^{2}\left(\frac{E}{E_{\rm QG,n}}\right)^{\rm n}\;,
\label{eq:dispersion}
\end{equation}
where $E_{\rm QG}$ is the QG energy scale, $m$ is the rest-mass of the particle,
the n-th order expansion of leading term corresponds to linear (n=1) or quadratic (n=2),
and $+1$ $(-1)$ stands for a decrease (increase) in particle speed with an increasing
particle energy (also referred to as the ``subluminal" and ``superluminal" cases).
The current best limits on the QG energy scale have been derived from the GeV photons of
GRB 090510. The limits set are $E_{\rm QG, 1}>9.1\times10^{19}$ GeV $>(1-10)E_{\rm Pl}$
and $E_{\rm QG, 2}>1.3\times10^{11}$ GeV for linear and quadratic violations of
Lorentz invariance, respectively \cite{Abdo et al.2009,Vasileiou et al.2013}
(see also \cite{Kostelecky et al.2011,Liberati2013} and summary constraints for Lorentz violation therein).
Ref. \cite{Ellis et al.2008} used the supernova 1987A neutrino data to study LIV effects,
and obtained the limits of $E_{\rm QG, 1}>2.7\times10^{10}$ GeV and $E_{\rm QG, 2}>4.6\times10^{4}$ GeV.
Based on the association between the outburst of of the blazar PKS B1424-418 and a PeV neutrino,
Ref. \cite{Wang et al.2016} set the most stringent limits up to now on neutrino LIV for subluminal neutrinos,
implying $E_{\rm QG, 1}>1.1\times10^{17}$ GeV and $E_{\rm QG, 2}>7.3\times10^{11}$ GeV.

Einstein's WEP is at the heart of general relativity as well as of other
metric theories of gravity. The WEP states that any two different types of massless (or negligible rest-mass) messenger particles,
or two of the same particles with different energies, if emitted simultaneously from the same
astrophysical object and passing through the same gravitational field,
should arrive at Earth at the same time \cite{Will2006,Will2014}. The measurements of the arrival times of
neutrinos and photons from supernova 1987A have been used \cite{Krauss et al.1988,Longo1988}
to constrain the violations of the WEP through the Shapiro (gravitational) time delay effect \cite{Shapiro1964}.
They showed that the Shapiro delays of neutrino and photon are equal to an accuracy of approximately
0.34\%. With the flight time difference between the PeV neutrino and the blazar photons, Ref. \cite{Wang et al.2016}
found that the constraints on the WEP accuracy from neutrinos can be further improved by two orders
of magnitude when taking into account the gravitational potential of clusters or superclusters.
Besides neutrinos, different-energy photons from extragalactic transients and variables have been applied to test the WEP
through the Shapiro time delay effect, such as the photon emissions from GRBs \cite{Gao et al.2015,Sivaram1999},
fast radio bursts \cite{Wei et al.2015,Nusser2016,Zhang2016,Tingay et al.2016}, and TeV blazars
\cite{Wei et al.2016}. Moreover, such a test can now be extended to include a new messenger,
namely gravitational waves \cite{Wu et al.2016,Kahya et al.2016}.

In this work, we determine the limits on the neutrino velocity,
Lorentz invariance, and the WEP that
would result if these five IceCube neutrinos were created in
the same events as the gamma-ray photons observed in
the associated GRBs.

\section{Observational data}
The IceCube Collaboration detected five neutrino events correlated with five GRBs \cite{IceCube Collaboration et al.2016}.
Taking this association at face value, the data of these events and GRBs are shown in
Table~\ref{table}. The first eight columns include the following information for each association:
(1) the source name; (2) the GRB prompt photon emission time $T_{100}$, which is defined by the most
inclusive start and end times ($T_{1}$ and $T_{2}$) reported by any satellite;
(3) the observed maximum time delay $\Delta t$ between the start time of the GRB
prompt photon emission and the arrival of the neutrino; the spatial parameters, including
(4) the right ascension coordinate, (5) the declination coordinate; (6) the observed gamma-ray fluence
of the GRBs; (7) the neutrino energy; and (8) the redshift.

The original idea of neutrino emission from GRBs \cite{Waxman et al.1997}
was that TeV neutrinos originate via interactions of shock-accelerated
protons with the observed $\gamma$ rays ($p\gamma$) in the same internal shocks which
produce the MeV photons, i.e., they would be contemporaneous with the
MeV photons. The same applies if the photons and neutrinos originate at the
photospheric surface, e.g. \cite{Gao et al.2012}. Within the astrophysical model uncertainties,
the difference between the proton and the electron acceleration times does not introduce
an appreciable difference as far as the arrival times of the observable secondaries.
The relativistic protons collide with photons produced by electrons
in the same region where both protons and electrons were accelerated and
where the photons were produced, so one does not expect a geometric delay.
A delay of the neutrinos could be due to the neutrinos being produced in the
sub-photosphere by ($pp$) and ($pn$) cascades, whose development would induce some
delay of order seconds (see e.g. \cite{Asano et al.2013}).
Another delay might be if the neutrinos originate from ($p\gamma$) photohadronic
processes in a different region than the MeV photons. E.g. the MeV photons
might come from internal shocks, or a photosphere, while the neutrinos might
come from the external reverse shock \cite{Waxman et al.2000}. This would
involve a geometric delay. However, in this case the neutrinos would be expected
to be the EeV ($10^{18}$ eV) range, since the protons would be interacting with
optical/ultraviolet photons, rather than with MeV photons. But the observed
neutrinos are TeV, so this is not a possibility.

In sum, it is reasonable to
assume that the TeV neutrinos originate at the same time at which the GRB photon
emission begins, or else that the neutrinos are observed of the order of seconds later
than the start time of the GRB prompt emission. It should be underlined that
our limits on fundamental physics are based on a relatively conservative estimate of
the observed time delay, since we use the time interval between the start time
of the GRB prompt emission and the arrival of the neutrino as the observed maximum time delay.

Except for GRB 101213A at $z=0.414$, the other four GRBs do not have measured redshifts.
The empirical luminosity relation (the so-called Amati relation \cite{Amati et al.2002})
is therefore applied to estimate the redshift range of the four GRBs. We use the
observed fluence, energy band, and spectral parameters of the four bursts\footnote{The relevant
observed fluence, energy band, and spectral parameters of GRBs are
available on the GRBweb database at http://grbweb.icecube.wisc.edu.}
to calculate the intrinsic isotropic gamma-ray energies and the intrinsic peak energies for different
redshifts.  By requiring that the bursts enter the $3\sigma$ region of the relation, we derive
$z\geq0.170$ for GRB 110101B, $z\geq0.237$ for GRB 110521B, $z\geq0.269$ for GRB 111212A,
and $z\geq0.197$ for GRB 120114A (the lower limits of redshifts are conservatively adopted
in the rest of this paper). Regarding the details on estimating the redshifts of GRBs,
see \cite{Deng et al.2014} for a more detailed description. Here we adopt the cosmological parameters
derived from the Planck observations \cite{Planck Collaboration et al.2014}:
$\Omega_{\rm m}=0.315$, $\Omega_{\Lambda}=0.685$, and $H_{0}=67.3$ km $\rm s^{-1}$ $\rm Mpc^{-1}$.

\begin{table*}
\caption{Limits on $|v-c|/c$, $E_{\rm QG, 1}$, $E_{\rm QG, 2}$, and $\Delta \gamma$
from the associations between the TeV neutrinos and GRBs.}
\tiny
\centering
\begin{tabular}{cccccccccccc}
\hline
\hline
 GRB/Neutrino &  $T_{100}$ & $\Delta t$ & $\rm R.A.^{a}$ & $\rm Dec.^{a}$ & Fluence & Energy & $z$ & $|v-c|/c$ & $E_{\rm QG, 1}$ & $E_{\rm QG, 2}$ & $\Delta \gamma$ \\
 & (s) & (s) & (deg) & (deg) & (erg cm$^{-2}$) & (TeV) &  &  & (GeV) & (GeV) & \\
\hline
101213A/Event 1	&	202 & 109	&	241.314	&	21.897	&$	7.4	\times10^{	-6	}$	&	11	&	0.414	&$	6.4	\times10^{	-16	}$	&$	2.1	\times10^{	19	}$	&$	 6.4	\times10^{	11	}$	&$	5.6	\times10^{	-11	}$	\\
110101B/Event 2	&	235 & 141	&	105.500	&	34.580	&$	6.6	\times10^{	-6	}$	&	34	&	$0.170^{b}$	&$	1.9	\times10^{	-15	}$	&$	2.0	\times10^{	19	}$	 &$	 1.0	\times10^{	12	}$	&$	9.6	\times10^{	-11	}$	 \\
110521B/Event 3	&	6.14 & 0.26	&	57.540	&	-62.340	&$	3.6	\times10^{	-6	}$	&	3.4	&	$0.237^{b}$	&$	2.5	\times10^{	-18	}$	&$	1.5	\times10^{	21	}$	 &$	2.9	\times10^{	12	}$	&$	1.3	\times10^{	-13	}$	\\
111212A/Event 4	&	68.5 & 11.7	&	310.431	&	-68.613	&$	1.4	\times10^{	-6	}$	&	30.6	&	$0.269^{b}$	&$	1.0	\times10^{	-16	}$	&$	3.4	\times10^{	20	 }$	&$	4.2	\times10^{	12	}$	&$	6.0	\times10^{	-12	}$ \\
120114A/Event 5	&	43.3 & 57.2	&	317.904	&	57.036	&$	2.4	\times10^{	-6	}$	&	3.8	&	$0.197^{b}$	&$	6.7	\times10^{	-16	}$	&$	6.3	\times10^{	18	}$	 &$	2.0	\times10^{	11	}$	&$	1.9	\times10^{	-11	}$	\\
\hline
\end{tabular}
\label{table}
\\
$^{\rm a}$The location information of GRBs are available on the GRBweb database at http://grbweb.icecube.wisc.edu.
\\
$^{\rm b}$The redshifts of the four GRBs are estimated by the luminosity relation.
\end{table*}

\section{New precision limits on fundamental physics with neutrinos from GRBs}

\subsection{Constraining the neutrino velocity}

Assuming the physical associations between the TeV neutrinos and GRBs, observational constraints on
the neutrino velocity can be obtained. The simplest parametrization of the constraint is in terms of
an effective limiting velocity of neutrinos $v$ compared to that of photons $c$. This limit on
the relative velocity difference will be \cite{Schaefer1999}
\begin{equation}
|v-c|/c\leq c\cdot\Delta t/D\;,
\end{equation}
where $D=\frac{c}{H_{0}}\int_{0}^{z}\frac{dz'}{\sqrt{\Omega_{m}(1+z')^{3}+\Omega_{\Lambda}}}$
is the co-moving distance of the source. The limits on $|v-c|/c$ for each event are presented in column 9 of
Table~\ref{table}. The strictest limit is $|v-c|/c\leq2.5\times10^{-18}$ for GRB 110521B/Event 3
and the worst limit is $1.9\times10^{-15}$ for GRB 110101B/Event 2, which are close to $10^{4}-10^{7}$ times
better than the neutrino-velocity limit obtained with the neutrino from
a blazar flare \cite{Kadler et al.2016}.

\subsection{Constraining violation of the Lorentz invariance}
Note that the superluminal neutrinos would loss their energy quickly
due to both vacuum pair emission and neutrino splitting \cite{Stecker et al.2015,Maccione et al.2013},
and some excellent limits on LIV have been derived from superluminal neutrinos
\cite{Borriello et al.2013,Diaz et al.2014,Stecker2014}. Here we set the limits
on LIV for subluminal neutrinos, i.e., the low energy photons travel faster than
the high energy neutrinos. As the test particles are massless or nearly massless,
the term $m^{2}c^{4}$ in Equation~(\ref{eq:dispersion}) is absolutely negligible.
Since the speed of particles have an energy dependence, two particles with different
energies emitted simultaneously from the source will arrive at the observer
with a time delay $\Delta t$. For a cosmic transient source, one has to consider the
cosmological expansion when calculating the LIV induced time delay (see e.g. 
\cite{Ellis et al.2008,Jacob et al.2008,Shao et al.2010,Chang et al.2012,Zhang et al.2015})
\begin{equation}
\Delta t=\frac{1+\rm n}{2H_{0}}\frac{E_{h}^{\rm n}-E_{l}^{\rm n}}{E_{\rm QG, n}^{\rm n}}
\int_{0}^{z}\frac{(1+z')^{\rm n}dz'}{\sqrt{\Omega_{m}(1+z')^{3}+\Omega_{\Lambda}}}\;,
\label{LIV}
\end{equation}
where $E_{h}$ and $E_{l}$ ($E_{h}>E_{l}$) correspond to the different particle energies.

The limits on $E_{\rm QG, 1}$ and $E_{\rm QG, 2}$ for each GRB neutrino are presented in columns 10 and 11
of Table~\ref{table}.\footnote{Note that $E_{h}^{\rm n}-E_{l}^{\rm n}$ in Equation~(\ref{LIV}) can be approximated as $E_{h}^{\rm n}$,
since the energies of GRB photons at the trigger time ($\sim 100$ keV) are several orders of magnitude lower
than those of neutrinos ($\sim$ TeV).}
The tightest limits on the linear and quadratic terms are
$E_{\rm QG, 1}>1.5\times10^{21}$ GeV$>100E_{\rm Pl}$ for GRB 110521B/Event 3 and
$E_{\rm QG, 2}>4.2\times10^{12}$ GeV for GRB 111212A/Event 4, respectively.
The worst limits are $E_{\rm QG, 1}>6.3\times10^{18}$ GeV and $E_{\rm QG, 2}>2.0\times10^{11}$ GeV
for GRB 120114A/Event 5.
We see that the linear ($\rm n=1$) LIV term can be easily excluded by GRB 110521B/Event 3.
Compared with the previous best limit on $E_{\rm QG, 1}$ from GeV photons from GRB 090510
($E_{\rm QG, 1}>9.1\times10^{19}$ GeV \cite{Abdo et al.2009}), our tightest limit on $E_{\rm QG, 1}$
represents an improvement of at least one order of magnitude, although our worst limit on
$E_{\rm QG, 1}$ is not as stringent as that of GRB 090510. Moreover, our limits on
$E_{\rm QG, 2}$ are as good as or even 10 times tighter than the previous best result from
the blazar PeV neutrino ($E_{\rm QG, 2}>7.3\times10^{11}$ GeV \cite{Wang et al.2016}).
In short, we set the most stringent limits on both the linear and quadratic LIV effects to date
under the assumption that the associations between the TeV neutrinos and GRBs would be finally confirmed.

Here we want to point out a possible caveat to the $\rm n=1$ limits.
In an effective field theory framework, the dispersion relation (see Equation~\ref{eq:dispersion})
for $\rm n=1$ would imply CPT odd terms which predict opposite signs
for the Lorentz breaking term for neutrino and antineutrino. In other words, if the
neutrino is e.g. superluminal, then the antineutrino will be subluminal.
However, the IceCube detector does not distinguish neutrinos from antineutrinos.
So for an initial equal amounts of neutrino and antineutrino, one expect that superluminal neutrinos will
probably decay in lower energy neutrinos and antineutrinos with the latter
having partially travel faster than expected. It is hard to know whether the two effects
can compensate each other or not. If not, this would add another error term to the observed time delay
and affect our purposed constraints at some level.

\subsection{Testing the weak equivalence principle}
Indeed, all metric theories of gravity satisfying the WEP predict that all test particles
must follow same trajectories and undergo the identical Shapiro time delay.
That is, as long as the WEP is valid, all metric theories expect $\gamma_{1}=\gamma_{2}\equiv\gamma$,
where $\gamma$ is the parametrized post-Newtonian (PPN) parameter and the subscripts
correspond to two different particles \cite{Will2006,Will2014,Jacobson et al.2001}. The WEP accuracy
can therefore be described by limits on the differences of $\gamma$ values for different particles.
According to the Shapiro time delay effect \cite{Shapiro1964},
the time interval for particles to traverse a given distance is longer by
\begin{equation}
t_{\rm gra}=-\frac{1+\gamma}{c^3}\int_{r_e}^{r_o}~U(r)dr
\end{equation}
in the presence of a gravitational potential $U(r)$, where the integration is along the path of
the particle emitted at $r_{e}$ and received at $r_{o}$.
If the values of the PPN parameter $\gamma$ are different for different particles,
two particles emitted simultaneously from the source will reach us at different times, and the corresponding
time delay is given by
\begin{equation}
\Delta t_{\rm gra}=\frac{\gamma_{\rm 1}-\gamma_{\rm 2}}{c^3}\int_{r_o}^{r_e}~U(r)dr\;,
\label{gra}
\end{equation}
where $\gamma_{\rm 1}-\gamma_{\rm 2}$ represents the difference between the $\gamma$ values for
different particles.

To calculate the Shapiro time delay with Equation~(\ref{gra}), we need to know
the gravitational potential $U(r)$. Most previous studies focused on the
contribution from the gravitational potential of the Milky Way. However,
it has been showed that incorporating the gravitational potential of the
large scale structure would tighten the constraints on the WEP accuracy
\cite{Nusser2016,Zhang2016,Wang et al.2016,Luo et al.2016}. We here consider
the gravitational potential of the Laniakea supercluster of galaxies.
Laniakea is a newly discovered supercluster of galaxies, in which the Milky Way
as well as the Local Group reside \cite{Tully et al.2014}. The total mass of Laniakea
is $10^{17}M_{\odot}$, which is about $10^{5}$ times heavier than the Milky Way.

Assuming that the measured time delays $(\Delta t)$ between correlated particles from the same source
are mainly caused by the gravitational potential of the Laniakea supercluster of galaxies,
and adopting a Keplerian potential for Laniakea\footnote{In fact,
       the gravitational potential of the Laniakea supercluster
       is still not well known. Besides the Keplerian potential model,
       Ref. \cite{Krauss et al.1988} also examined the other widely used potential
       model, i.e., the isothermal potential. They showed that these two different
       potential models do not have a strong influence on the results for
       testing the WEP. Here we adopt the Keplerian potential for
       the whole supercluster.},
we have \cite{Longo1988}
\begin{eqnarray}
\Delta t>\Delta t_{\rm gra}= \left(\gamma_{1}-\gamma_{2}\right ) \frac{GM_{\rm L}}{c^{3}} \times \qquad\qquad\qquad\qquad\qquad\\ \nonumber
\ln \left\{ \frac{ \left[d+\left(d^{2}-b^{2}\right)^{1/2}\right] \left[r_{L}+s_{\rm n}\left(r_{L}^{2}-b^{2}\right)^{1/2}\right] }{b^{2}} \right\}\;,
\label{eq:gammadiff}
\end{eqnarray}
where $M_{\rm L}\simeq10^{17}M_{\odot}$ is the Laniakea mass \cite{Tully et al.2014},
$d$ denotes the distance from the source to the Laniakea center (for a cosmic source,
$d$ is approximated as the distance from the source to the Earth), $b$ represents
the impact parameter of the particle paths relative to the Laniakea center,
and $s_{\rm n}=\pm1$ corresponds to the sign of the correction of the source direction.
If $s_{\rm n}=+1$ ($s_{\rm n}=-1$), the source is located along the direction of Laniakea (anti-Laniakea) center.
Since the gravitational center of Laniakea is considered to be the Great Attractor \cite{Lynden-Bell et al.1988},
a mass concentration in the nearby Universe, we adopt the coordinates of the Great Attractor center
($\rm R.A.=10^{h}32^{m}$, $\rm Dec.=-46^{\circ}00^{'}$) instead of that of the Laniakea center.
For a cosmic source in the position ($\rm R.A.=\beta_{s}$, $\rm Dec.=\delta_{s}$),
the impact parameter can be written as
\begin{equation}
b=r_{L}\sqrt{1-(\sin \delta_{s} \sin \delta_{L}+\cos \delta_{s} \cos \delta_{L} \cos(\beta_{s}-\beta_{L}))^{2}}\;,
\end{equation}
where $r_{L}=79$ Mpc is the distance from the Earth to the center of Laniakea, and
($\beta_{L}=10^{\rm h}32^{\rm m}$, $\delta_{L}=-46^{\circ}00^{'}$)
represent the coordinates of the Laniakea center.

Thus, with the observed time delays between the beginning of the GRB prompt photon emission and
the arrival of the neutrinos in Table~\ref{table}, we obtain WEP constraints from Equation~(\ref{eq:gammadiff})
for assuming that these five TeV neutrinos are truly associated with GRBs.
The limits on $\gamma_{\nu}-\gamma_{\gamma}$ for each event are displayed in column 12 of Table~\ref{table}.
The strictest limit is $\gamma_{\nu}-\gamma_{\gamma}<1.3\times10^{-13}$ for GRB 110521B/Event 3
and the worst limit is $9.6\times10^{-11}$ for GRB 110101B/Event 2,
which are about $5-7$ orders of magnitude more constraining than the previous limit obtained with the
PeV neutrino from a blazar flare ($\gamma_{\nu}-\gamma_{\gamma}<7.0\times10^{-6}$ \cite{Wang et al.2016}).

\begin{table*}
\tiny
\centering
\caption{Limits on $|v-c|/c$, $E_{\rm QG, 1}$, $E_{\rm QG, 2}$, and $\Delta \gamma$
using the GRB durations $T_{100}$ as the conservative limits of the observed time delays.}
\begin{tabular}{cccccccccccc}
\hline
\hline
 GRB/Neutrino &  $T_{100}$ & $\Delta t$ & $\rm R.A.$ & $\rm Dec.$ & Fluence & Energy & $z$ & $|v-c|/c$ & $E_{\rm QG, 1}$ & $E_{\rm QG, 2}$ & $\Delta \gamma$ \\
 & (s) & (s) & (deg) & (deg) & (erg cm$^{-2}$) & (TeV) &  &  & (GeV) & (GeV) & \\
\hline
101213A/Event 1	&	202 & 109	&	241.314	&	21.897	&$	7.4	\times10^{	-6	}$	&	11	&	0.414	&$	1.2	\times10^{	-15	}$	&$	1.1	\times10^{	19	}$	&$	 4.7	\times10^{	11	}$	&$	1.0	\times10^{	-10	}$	\\
110101B/Event 2	&	235 & 141	&	105.500	&	34.580	&$	6.6	\times10^{	-6	}$	&	34	&	0.170	&$	3.1	\times10^{	-15	}$	&$	1.2	\times10^{	19	}$	&$	 8.1	\times10^{	11	}$	&$	1.6	\times10^{	-10	}$	\\
110521B/Event 3	&	6.14 & 0.26	&	57.540	&	-62.340	&$	3.6	\times10^{	-6	}$	&	3.4	&	0.237	&$	6.0	\times10^{	-17	}$	&$	6.3	\times10^{	19	}$	 &$	6.0	\times10^{	11	}$	&$	3.0	\times10^{	-12	}$	\\
111212A/Event 4	&	68.5 & 11.7	&	310.431	&	-68.613	&$	1.4	\times10^{	-6	}$	&	30.6	&	0.269	&$	5.9	\times10^{	-16	}$	&$	5.8	\times10^{	19	 }$	&$	1.7	\times10^{	12	}$	&$	3.5	\times10^{	-11	}$ \\
120114A/Event $5^{a}$	&	43.3 & 57.2	&	317.904	&	57.036	&$	2.4	\times10^{	-6	}$	&	3.8	&	0.197	&$	6.7	\times10^{	-16	}$	&$	6.3	\times10^{	18	 }$	 &$	2.0	\times10^{	11	}$	&$	1.9	\times10^{	-11	}$	 \\
\hline
\end{tabular}
\label{table2}
\\
$^{\rm a}$Note that IceCube Collaboration et al. (2016) accept neutrino
events out to $T_{100}\pm4\sigma_{t}$ for each GRB time window, where $\sigma_{t}$ is the width of
the Gaussian tails before $T_{1}$ and after $T_{2}$. Hence, it is reasonable to have $\Delta t>T_{100}$
for GRB 120114A/Event 5. For this association, here we still adopt the value of $\Delta t$ as the observed time delay.
\end{table*}

\section{Summary and discussion}

In this work, we discuss the potential of five possible GRB/neutrino associations from the
IceCube neutrino observatory for probing fundamental physics. We show that if the GRB identifications
are verified, significant improvements can be obtained on limits on the neutrino velocity, the
violation of Lorentz invariance, and the accuracy of the WEP, by using the observed time delays
between the neutrinos and photons. First,
the strictest limit on the relative velocity difference is $|v-c|/c\leq2.5\times10^{-18}$ for
GRB 11052B/Event 3, and the worst limit still comes to $1.9\times10^{-15}$ for GRB 110101B/Event 2,
which are about $10^{4}-10^{7}$ times better than the neutrino-velocity limit obtained with
the neutrino from a blazar flare. Secondly, we place the most stringent limits to date on
both the linear and quadratic LIV terms, namely $E_{\rm QG, 1}>6.3\times10^{18}-1.5\times10^{21}$ GeV
and $E_{\rm QG, 2}>2.0\times10^{11}-4.2\times10^{12}$ GeV, which are as good as or are an improvement
of one order of magnitude over the results previously obtained from the GeV photons of GRB 090510 and
the blazar PeV neutrino. Finally, we find that the limits on the differences of the $\gamma$ values
for the neutrinos and photons are as low as $\sim 10^{-11}-10^{-13}$, improving the limits on the WEP
from the blazar/neutrino association by at least $5-7$ orders of magnitude.
Nevertheless, it should be underlined that these limits are at best forecast
of what could be achieved if the GRB/neutrino associations would be finally verified.

As described above, for a GRB/neutrino association, the neutrinos would be contemporaneous
with the MeV photons. We have first set a relatively conservative limit by using the time interval
between the start time
of the GRB prompt emission and the arrival of the neutrino as the observed time delay when testing
fundamental physics. However, the neutrino and photon production is stochastic (and much fewer
neutrinos are produced than photons), so the first neutrino could have been produced well after
the first photon (but before the last photon, if the same shocks that accelerate electrons
radiating photons are also responsible for accelerating the protons which produce neutrinos).
That is, the observed time delay of any neutrino and photon
in the association system couldn't be longer than the GRB prompt photon emission time ($T_{100}$).
To account for the uncertainty of the observed time delay, we also test one more case by assuming
$T_{100}$ as the conservative limit of the observed time delay. The more conservative limits on
$|v-c|/c$, $E_{\rm QG, 1}$, $E_{\rm QG, 2}$, and $\Delta \gamma$ are shown in Table~\ref{table2}.
One can see from this table that the implications of the GRB neutrino tests of fundamental physics are not greatly affected.
Compared with those constraints presented in Table~\ref{table}, the new constraint results only vary within a factor
of two to 1 order of magnitude.

It has long been proposed that high energy neutrinos would be associated with GRBs.
Thanks to the great wide field of view and high sensitivity,
very-large-volume neutrino telescopes such as ANTARES and IceCube
are ideal detectors to search for any high-energy
neutrinos from GRBs, e.g. \cite{Adrian-Martinez et al.2013,Aartsen et al.2014}.
Although the dedicated searches for high energy neutrinos correlated with GRBs have led to null results
\cite{Abbasi et al.2012,Aartsen et al.2015}, it is entirely possible that some of the IceCube neutrinos
originate from GRBs, such as these five spatial and temporal coincidences for the neutrinos and GRBs
\cite{IceCube Collaboration et al.2016}. If in the future the origin of these IceCube neutrinos is
better understood, or the associations between different detected events are truly confirmed,
the prospects for testing fundamental physics with high energy cosmic neutrinos,
as discussed in this work, will be very promising.

\acknowledgments
We are grateful to the anonymous referee for the helpful suggestions
which have helped us to improve the presentation of the paper.
We also thank Lijing Shao and Xiang-Yu Wang for the useful communications.
This work is partially supported by the National Basic Research Program (``973'' Program)
of China (Grants 2014CB845800 and 2013CB834900), the National Natural Science Foundation
of China (grants Nos. 11322328, 11433009, and 11543005), the Natural Science Foundation
of Jiangsu Province (Grant No. BK20161096), the Youth Innovation Promotion Association (2011231),
the Strategic Priority Research Program ``The Emergence of Cosmological Structures'' (Grant No. XDB09000000) of
the Chinese Academy of Sciences, the Guangxi Key Laboratory for Relativistic Astrophysics, and NASA NNX13AH50G.


\providecommand{\href}[2]{#2}\begingroup\raggedright\begin{thebibliography}{}

\end{thebibliography}\endgroup


\begin{thebibliography}{99}

\bibitem{IceCube Collaboration et al.2016} IceCube Collaboration,
M.~G.~Aartsen, K.~Abraham, M.~Ackermann, J.~Adams, J.~A.~Aguilar, et al.,
{\it An All-Sky Search for Three Flavors of Neutrinos from Gamma-Ray Bursts
with the IceCube Neutrino Observatory},
arXiv:1601.06484.

\bibitem{Jacob et al.2007} Jacob,
U.~\& T.~Piran,  {\it Neutrinos from gamma-ray bursts as a tool to explore quantum-gravity-induced Lorentz violation}, {\it Nature Physics}
{\bf 3} (2007) 87,  [arXiv:hep-ph/0607145].

\bibitem{Gao et al.2015} Gao, H., X.-F.~Wu,
\& P.~M{\'e}sz{\'a}ros,  {\it Cosmic Transients Test Einstein's Equivalence Principle out to GeV Energies}, {\it ApJ} {\bf 810} (2015) 121, [arXiv:1509.00150].

\bibitem{Wang et al.2016} Wang, Z.-Y., R.-Y.~Liu,
\& X.-Y.~Wang,  {\it Testing the Equivalence Principle and Lorentz Invariance with PeV Neutrinos from Blazar Flares}, {\it Phys. Rev. Lett.} {\bf 116} (2016) 151101,  [arXiv:1602.06805].

\bibitem{Stecker et al.2015} Stecker, F.~W., S.~T.~Scully, S.~Liberati,
\& D.~Mattingly,  {\it Searching for traces of Planck-scale physics with high energy neutrinos}, {\it Phys. Rev. D} {\bf 91} (2015) 045009,
[arXiv:1411.5889].

\bibitem{Longo1987} Longo, M.~J.,  {\it Tests of relativity from SN1987A},
{\it Phys. Rev. D} {\bf 36} (1987) 3276.

\bibitem{Kadler et al.2016} Kadler, M., F.~Krau{\ss}, K.~Mannheim, R.~Ojha,
C.~M{\"u}ller, R.~Schulz, et al.,  {\it Coincidence of a high-fluence
blazar outburst with a PeV-energy neutrino event}, {\it Nature Physics}, in press,
arXiv:1602.02012.

\bibitem{Amelino-Camelia et al.1997} Amelino-Camelia, G., J.~Ellis,
N.~E.~Mavromatos,
\& D.~V.~Nanopoulos,  {\it Distance Measurement and Wave Dispersion in a Liouville-String Approach to Quantum Gravity}, {\it International
Journal of Modern Physics A} {\bf 12} (1997) 607,  [arXiv:hep-th/9605211].

\bibitem{Amelino-Camelia et al.1998} Amelino-Camelia, G., J.~Ellis,
N.~E.~Mavromatos, D.~V.~Nanopoulos,
\& S.~Sarkar,  {\it Tests of quantum gravity from observations of $\gamma$-ray bursts}, {\it Nature} {\bf 393} (1998) 763,  [arXiv:astro-
ph/9712103].

\bibitem{Mattingly2005} Mattingly, D.,  {\it Modern Tests of Lorentz
Invariance}, {\it Living Reviews in Relativity} {\bf 8} (2005)
[arXiv:gr-qc/0502097].

\bibitem{Amelino-Camelia2013} Amelino-Camelia, G.,  {\it Quantum-Spacetime
Phenomenology}, {\it Living Reviews in Relativity} {\bf 16} (2013)
[arXiv:0806.0339].

\bibitem{Choubey et al.2004} Choubey,
S.~\& S.~F.~King,  {\it Electrophobic Lorentz invariance violation for neutrinos and the see-saw mechanism}, {\it Phys. Lett. B} {\bf 586}
(2004) 353,  [arXiv:hep-ph/0311326].

\bibitem{Ellis et al.2013} Ellis,
J.~\& N.~E.~Mavromatos,  {\it Probes of Lorentz violation}, {\it Astroparticle Physics} {\bf 43} (2013) 50,  [arXiv:1111.1178].

\bibitem{Abdo et al.2009} Abdo, A.~A., M.~Ackermann, M.~Ajello, K.~Asano,
W.~B.~Atwood, M.~Axelsson, et al.,  {\it A limit on the variation of the
speed of light arising from quantum gravity effects}, {\it Nature} {\bf 462}
(2009) 331,  [arXiv:0908.1832].

\bibitem{Vasileiou et al.2013} Vasileiou, V., A.~Jacholkowska, F.~Piron,
J.~Bolmont, C.~Couturier, J.~Granot, et al.,  {\it Constraints on Lorentz
invariance violation from Fermi-Large Area Telescope observations of
gamma-ray bursts}, {\it Phys. Rev. D} {\bf 87} (2013) 122001,  [arXiv:1305.3463].

\bibitem{Kostelecky et al.2011} Kosteleck{\'y},
V.~A.~\& N.~Russell,  {\it Data tables for Lorentz and CPT violation}, {\it Reviews of Modern Physics} {\bf 83} (2011) 11,  [arXiv:0801.0287].

\bibitem{Liberati2013} Liberati, S.,  {\it Tests of Lorentz invariance: a
2013 update}, {\it Classical and Quantum Gravity} {\bf 30} (2013) 133001,
[arXiv:1304.5795].

\bibitem{Ellis et al.2008} Ellis, J., N.~Harries, A.~Meregaglia, A.~Rubbia,
\& A.~S.~Sakharov,  {\it Probes of Lorentz violation in neutrino propagation}, {\it Phys. Rev. D} {\bf 78} (2008) 033013,  [arXiv:0805.0253].

\bibitem{Will2006} Will, C.~M.,  {\it The Confrontation between General
Relativity and Experiment}, {\it Living Reviews in Relativity} {\bf 9}
(2006)  [arXiv:gr-qc/0510072].

\bibitem{Will2014} Will, C.~M.,  {\it The Confrontation between General
Relativity and Experiment}, {\it Living Reviews in Relativity} {\bf 17}
(2014)  [arXiv:1403.7377].

\bibitem{Krauss et al.1988} Krauss,
L.~M.~\& S.~Tremaine,  {\it Test of the weak equivalence principle for neutrinos and photons}, {\it Phys. Rev. Lett.} {\bf 60} (1988)
176.

\bibitem{Longo1988} Longo, M.~J.,  {\it New precision tests of the Einstein
equivalence principle from SN1987A}, {\it Phys. Rev. Lett.} {\bf 60}
(1988) 173.

\bibitem{Shapiro1964} Shapiro, I.~I.,  {\it Fourth Test of General
Relativity}, {\it Phys. Rev.. Lett.} {\bf 13} (1964) 789.

\bibitem{Sivaram1999} Sivaram, C.,  {\it Constraints on the photon mass and
charge and test of equivalence principle from GRB 990123}, {\it Bulletin of
the Astronomical Society of India} {\bf 27} (1999) 627.

\bibitem{Wei et al.2015} Wei, J.-J., H.~Gao, X.-F.~Wu,
\& P.~M{\'e}sz{\'a}ros,  {\it Testing Einstein's Equivalence Principle With Fast Radio Bursts}, {\it Phys. Rev. Lett.} {\bf 115} (2015)
261101,  [arXiv:1512.07670].

\bibitem{Nusser2016} Nusser, A.,  {\it On Testing the Equivalence Principle
with Extragalactic Bursts}, {\it ApJL} {\bf 821} (2016) L2,
[arXiv:1601.03636].

\bibitem{Zhang2016} Zhang, S.-N.,  {\it Testing Einstein's Equivalence
Principle with Cosmological Fast Radio Bursts behind Clusters of Galaxies},
arXiv:1601.04558.

\bibitem{Tingay et al.2016} Tingay,
S.~J.~\& D.~L.~Kaplan,  {\it Limits on Einstein's Equivalence Principle from the First Localized Fast Radio Burst FRB 150418}, {\it
ApJL} {\bf 820} (2016) L31,  [arXiv:1602.07643].

\bibitem{Wei et al.2016} Wei, J.-J., J.-S.~Wang, H.~Gao,
\& X.-F.~Wu,  {\it Tests of the Einstein Equivalence Principle Using TeV Blazars}, {\it ApJL} {\bf 818} (2016) L2,  [arXiv:1601.04145].

\bibitem{Wu et al.2016} Wu, X.-F., H.~Gao, J.-J.~Wei,
P.~M{\'e}sz{\'a}ros, B.~Zhang, et al.,  {\it Testing Einstein's Weak Equivalence
Principle With Gravitational Waves}, {\it Phys. Rev. D} {\bf 94}
(2016) 024061, [arXiv:1602.01566].

\bibitem{Kahya et al.2016} Kahya,
E.~O.~\& S.~Desai,  {\it Constraints on frequency-dependent violations of Shapiro delay from GW150914}, {\it Phys. Lett. B} {\bf 756}
(2016) 265,  [arXiv:1602.04779].

\bibitem{Waxman et al.1997} Waxman,
E.~\& J.~Bahcall,  {\it High Energy Neutrinos from Cosmological Gamma-Ray Burst Fireballs}, {\it Phys. Rev. Lett.} {\bf 78} (1997) 2292,
[arXiv:astro-ph/9701231].

\bibitem{Gao et al.2012} Gao, S., K.~Asano,
\& P.~M{\'e}sz{\'a}ros,  {\it High energy neutrinos from dissipative photospheric models of gamma ray bursts}, {\it JCAP} {\bf 11} (2012) 058,
 [arXiv:1210.1186].

\bibitem{Asano et al.2013} Asano,
K.~\& P.~M{\'e}sz{\'a}ros,  {\it Photon and neutrino spectra of time-dependent photospheric models of gamma-ray bursts}, {\it JCAP} {\bf 9}
(2013) 008,  [arXiv:1308.3563].

\bibitem{Waxman et al.2000} Waxman,
E.~\& J.~N.~Bahcall,  {\it Neutrino Afterglow from Gamma-Ray Bursts: $\sim10^{18}$ EV}, {\it ApJ} {\bf 541} (2000) 707,  [arXiv:hep-
ph/9909286].

\bibitem{Amati et al.2002} Amati, L., F.~Frontera, M.~Tavani, J.~J.~M.~in't
Zand, A.~Antonelli, E.~Costa, et al.,  {\it Intrinsic spectra and
energetics of BeppoSAX Gamma-Ray Bursts with known redshifts}, {\it A\&A}
{\bf 390} (2002) 81,  [arXiv:astro-ph/0205230].

\bibitem{Deng et al.2014} Deng,
W.~\& B.~Zhang,  {\it Cosmological Implications of Fast Radio Burst/Gamma-Ray Burst Associations}, {\it ApJL} {\bf 783} (2014) L35,
[arXiv:1401.0059].

\bibitem{Planck Collaboration et al.2014} Planck Collaboration,
P.~A.~R.~Ade, N.~Aghanim, C.~Armitage-Caplan, M.~Arnaud, M.~Ashdown, et
al.,  {\it Planck 2013 results. XVI. Cosmological parameters}, {\it A\&A}
{\bf 571} (2014) A16,  [arXiv:1303.5076].

\bibitem{Schaefer1999} Schaefer, B.~E.,  {\it Severe Limits on Variations
of the Speed of Light with Frequency}, {\it Phys. Rev. Lett.} {\bf
82} (1999) 4964,  [arXiv:astro-ph/9810479].

\bibitem{Maccione et al.2013} Maccione, L., S.~Liberati,
\& D.~M.~Mattingly,  {\it Violations of Lorentz invariance in the neutrino sector: an improved analysis of anomalous threshold constraints}, {\it JCAP} {\bf 3} (2013) 039. 

\bibitem{Borriello et al.2013} Borriello, E., S.~Chakraborty, A.~Mirizzi,
\& P.~D.~Serpico,  {\it Stringent constraint on neutrino Lorentz invariance violation from the two IceCube PeV neutrinos}, {\it Phys. Rev. D} {\bf 87}
(2013) 116009,  [arXiv:1303.5843].

\bibitem{Diaz et al.2014} D{\'{\i}}az, J.~S., V.~A.~Kosteleck{\'y},
\& M.~Mewes,  {\it Testing relativity with high-energy astrophysical neutrinos}, {\it Phys. Rev. D} {\bf 89} (2014) 043005,  [arXiv:1308.6344].

\bibitem{Stecker2014} Stecker, F.~W.,  {\it Limiting superluminal electron
and neutrino velocities using the 2010 Crab Nebula flare and the IceCube
PeV neutrino events}, {\it Astroparticle Physics} {\bf 56} (2014) 16,
[arXiv:1306.6095].

\bibitem{Jacob et al.2008} Jacob,
U.~\& T.~Piran,  {\it Lorentz-violation-induced arrival delays of cosmological particles}, {\it JCAP} {\bf 1} (2008) 031,  [arXiv:0712.2170].

\bibitem{Shao et al.2010} Shao, L., Z.~Xiao,
\& B.-Q.~Ma,  {\it Lorentz violation from cosmological objects with very high energy photon emissions}, {\it Astroparticle Physics} {\bf 33}
(2010) 312,  [arXiv:0911.2276].

\bibitem{Chang et al.2012} Chang, Z., Y.~Jiang,
\& H.-N.~Lin,  {\it A unified constraint on the Lorentz invariance violation from both short and long GRBs}, {\it Astroparticle Physics} {\bf
36} (2012) 47,  [arXiv:1201.3413].

\bibitem{Zhang et al.2015} Zhang,
S.~\& B.-Q.~Ma,  {\it Lorentz violation from gamma-ray bursts}, {\it Astroparticle Physics} {\bf 61} (2015) 108,  [arXiv:1406.4568].

\bibitem{Jacobson et al.2001} Jacobson, T., \& D.~Mattingly, {\it Gravity with a dynamical preferred frame}, {\it Phys. Rev. D} {\bf 64} (2001) 024028,  [arXiv:gr-qc/0007031].

\bibitem{Luo et al.2016} Luo, Z.-X., B.~Zhang, J.-J.~Wei,
\& X.-F.~Wu,  {\it Testing Einstein's Equivalence Principle with supercluster Laniakea's gravitational field}, {\it Journal of High Energy
Astrophysics} {\bf 9} (2016) 35,  [arXiv:1604.02566].

\bibitem{Tully et al.2014} Tully, R.~B., H.~Courtois, Y.~Hoffman,
\& D.~Pomar{\`e}de,  {\it The Laniakea supercluster of galaxies}, {\it Nature} {\bf 513} (2014) 71,  [arXiv:1409.0880].

\bibitem{Lynden-Bell et al.1988} Lynden-Bell, D., S.~M.~Faber, D.~Burstein,
R.~L.~Davies, A.~Dressler, R.~J.~Terlevich, et al.,  {\it Spectroscopy and
photometry of elliptical galaxies. V - Galaxy streaming toward the new
supergalactic center}, {\it ApJ} {\bf 326} (1988) 19.


\bibitem{Adrian-Martinez et al.2013} Adri{\'a}n-Mart{\'{\i}}nez,
S., A.~Albert, I.~A.~Samarai, M.~Andr{\'e}, M.~Anghinolfi, G.~Anton, et
al.,  {\it Search for muon neutrinos from gamma-ray bursts with the ANTARES
neutrino telescope using 2008 to 2011 data}, {\it A\&A} {\bf 559} (2013)
A9,  [arXiv:1307.0304].

\bibitem{Aartsen et al.2014} Aartsen, M.~G., M.~Ackermann, J.~Adams,
J.~A.~Aguilar, M.~Ahlers, M.~Ahrens, et al.,  {\it Observation of
High-Energy Astrophysical Neutrinos in Three Years of IceCube Data}, {\it
Phys. Rev. Lett.} {\bf 113} (2014) 101101,  [arXiv:1405.5303].

\bibitem{Abbasi et al.2012} Abbasi, R., Y.~Abdou, T.~Abu-Zayyad,
M.~Ackermann, J.~Adams, J.~A.~Aguilar, et al.,  {\it An absence of
neutrinos associated with cosmic-ray acceleration in $\gamma$-ray bursts}, {\it
Nature} {\bf 484} (2012) 351,  [arXiv:1204.4219].

\bibitem{Aartsen et al.2015} Aartsen, M.~G., M.~Ackermann, J.~Adams,
J.~A.~Aguilar, M.~Ahlers, M.~Ahrens, et al.,  {\it Search for Prompt
Neutrino Emission from Gamma-Ray Bursts with IceCube}, {\it ApJL} {\bf
805} (2015) L5,  [arXiv:1412.6510].


\end{thebibliography}
\end{document}